# Interaction of nanoparticle properties and X-ray analytical techniques

Rainer Unterumsberger[1*], Philipp Hönicke[1], Yves Kayser[1], Beatrix Pollakowski-Herrmann[1], Saeed Gholhaki[2], Quanmin Guo[2], Richard E. Palmer[3] and Burkhard Beckhoff[1]

1: Physikalisch-Technische Bundesanstalt, Abbestraße 2-12, 10587 Berlin, Germany
2: School of Physics and Astronomy, University of Birmingham, Birmingham, UK B15 2TT, UK
3: College of Engineering, Swansea University, Bay Campus, Fabian Way, Swansea, SA1 8EN, UK

**Abstract**

In this work, Pt-Ti core-shell nanoparticles (NP) of 2 nm to 3 nm in size and 30000 u $\pm$ 1500 u as specified single particle mass, deposited on flat silicon substrates by means of a mass-selected cluster beam source, were used for the investigation of the modification of the X-Ray Standing Wave (XSW) field intensity with increasing NP surface coverage. The focus of the investigation is on the determination of the range of validity of the undisturbed flat surface approach of the XSW intensity in dependence of the actual coverage rate of the surface. Therefore, the nanoparticles were characterized using reference-free grazing incidence X-ray fluorescence analysis (GIXRF) employing radiometrically calibrated instrumentation. In addition, near-edge X-ray absorption fine structure (NEXAFS) measurements were performed to investigate the binding state of titanium in the core-shell nanoparticles which was found to be amorphous $TiO_2$. The combination of GIXRF measurements and of the calculated XSW field intensities allow for a quantification of the core-shell nanoparticle surface coverage. For six different samples, the peak surface coverage could be determined to vary from 7 % to 130 % of a complete monolayer-equivalent coverage. A result of the current investigation is that core-shell nanoparticles modify the intensity distribution of the XSW field with increasing surface coverage. This experimental result is in line with calculated XSW field intensity distributions at different surface coverages using an effective density approach.

## 1. Introduction

The development of the design and synthesis of nanoparticles is proceeding rapidly [1, 2] and the applications for nanoparticles reach more and more fields. They are being used in a wide range of applications such as in the industrial production of e.g. batteries for energy storage [3], included in everyday materials such as cosmetics or clothes [4, 5], life science and modern medicine e.g. to deliver diagnostics and treatments [6,7] or used as catalysts [8]. Recent developments allow for the production of precisely size-controlled nanoparticles [6]. The commonly used method for the characterization of such "atomic clusters" is the analysis of scanning transmission electron microscope (STEM) images [9]. For the specific Pt-Ti nanoparticle, this method has been used before to identify the Pt atoms and calculate the number of atoms from the determined volume of a Pt sphere [10]. The stability and catalytic properties of nanoparticles may be improved by combing two or more metallic elements [11, 12, 13]. Theoretical calculations suggest that the addition of Ti to Pt nanoparticle minimizes the catalytic poisoning of Pt by improving the reaction kinetics in Polymer Electrolyte Membrane (PEM) fuel cells [14, 15]. In addition, it has been shown that the addition of Pt atoms improves the photoactivity of $TiO_2$ by enhancing the separation between the photo-formed electrons and holes [16].

---

[*] Corresponding author

Nanoparticles have been investigated with XRF methods in different geometries such as total reflection (TXRF) [17, 18], grazing incidence (GIXRF) [19] or grazing exit (GEXRF) [20]. These methods provide element specific, non-destructive and non-preparative investigations. It has been shown that the morphology (average shape and size) of nanoparticles can be determined with these methods [20] and that the coverage of the substrate surface can modify the XSW field intensity [19, 21]. A surface coverage of 100 % means that a monolayer consisting of nanoparticles is covering the unit area. In this work, both aspects have been combined. In the first step, complementary information about very monodispersed Pt-Ti core-shell nanoparticles with respect to their mass deposition (being the mass per unit area of a specific element of the NPs), surface coverage and chemical binding state using SI-traceable X-ray spectrometry (XRS) [22, 23, 24, 25] methods are provided. The main part of the characterization is to determine the mass deposition and surface coverage using reference-free XRF. In addition, the advantage of having highly monodisperse nanoparticles is also the ability to investigate the strength and limits of the XSW field calculation required for quantitative GIXRF analysis.

## 2. Experimental approach

At the Physikalisch-Technische Bundesanstalt (PTB), the national metrology institute of Germany, the working group "X-ray Spectrometry" is determining elemental mass depositions by means of reference-free, fundamental parameter based X-ray fluorescence analysis. In the PTB laboratory at the electron storage ring BESSY II [26], radiometrically calibrated instrumentation is being operated. For the characterization of surface contamination, thin layers or nanoparticles using the reference-free XRF method, well-known experimental and instrumental parameters are required to ensure the traceability to the SI. The measurements were performed at the plane grating monochromator (PGM) beamline [22, 27, 28] for undulator radiation, the four-crystal monochromator beamline (FCM) [29] as well as at the 7-T wavelength shifter (WLS) beamline, the BAMline [30, 31]. All beamlines provide tunable, monochromatic synchrotron radiation with both high spectral purity and photon flux.

### 2.1. Monodispersed Nanoparticle samples

The samples were prepared in the University of Birmingham (BHAM) [10]. The core-shell nanoparticles were produced using a specialized magnetron sputtering technique [32] with 75 % titanium and 25 % platinum by weight in the sputtering target, corresponding to 13 Ti atoms for every Pt atom. For a detailed description of the nanoparticles and their production, we refer to Blackmore et al. [10]. They consist of nominal Pt-$TiO_2$ nanoparticles in a core-shell morphology. The core is made of Pt and the shell is made of $TiO_2$. The nominal NP surface coverages are 2.5 %, 5 %, 7.5 %, 15 %, 30 % and 60 %, deposited on flat Si substrates with native oxide top layer. Those coverage values were calculated by measuring the cluster current during deposition. In section 3.2, the determination of the surface coverage is described in details. Figure 1 shows the High Angle Annular Dark Field (HAADF) [33, 34] signal from a STEM image of the core-shell nanoparticles deposited on copper TEM grids covered with amorphous carbon film. The single particle mass of the nanoparticle is specified to be 30000 u $\pm$ 1500 u for Pt/Ti utilizing the mass filter in the cluster beam deposition source [35, 36], without considering possible oxidation processes, which occur when the samples were transferred from the cluster source to the STEM. The deposited area is about 1 $cm^2$; the cluster beam was moved in a raster scan during deposition to produce a uniform surface coverage. The nominal size of a core-shell nanoparticle is about 2 nm to 3 nm (determined using STEM image analysis [6]). Since the contrast in the HAADF signal from the STEM image is rather small for the titanium atoms, the uncertainty of the size-determination from the total nanoparticles is increased in comparison to the size-determination of the Pt-core. By determining the elemental composition of the sample and by means of additional knowledge of the total mass of one nanoparticle, the size can be determined using XRF methods (see section 3.1). Thus, analytical nanometrology (XRF) and dimensional nanometrology (STEM) can be combined. Without the knowledge of the mass deposition per core-shell nanoparticle, the absolute mass deposition of titanium cannot be transferred into a surface coverage (see section 3.2).

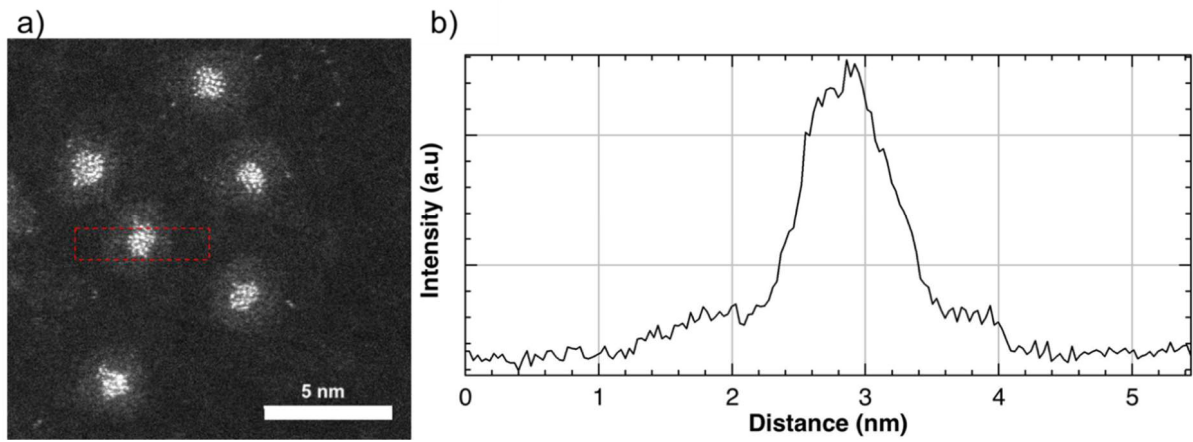

*Figure 1: (a) HAADF STEM image of the 30000 u core-shell nanoparticles. The bright cores in the middle are Pt and the halos around these cores are Ti oxidized by exposure to ambient on the transfer from the cluster source to the STEM, electron beam energy was 200 kV. (b) Averaged line profile of the dotted red rectangle on the nanoparticles, the shoulders demonstrate the intensity of $TiO_2$ on carbon and the peak represent combined Pt and $TiO_2$ intensity on carbon film.*

## 2.2. XRF and GIXRF measurements

For the determination of the elemental composition of the core-shell nanoparticles, XRF measurements were performed in an ultra-high vacuum chamber [37] optimized for reference-free XRF in various geometries. The samples can be aligned using translation and rotation motors to the pivot point of the chamber with respect to the incident excitation radiation, in order to irradiate all samples at their respective centers for all incident angles. The flux of the excitation radiation is detected with radiometrically calibrated photodiodes with a relative uncertainty in the responsivity of 1.0 % [29]. The emitted fluorescence radiation is detected by means of a silicon drift detector (SDD), which is calibrated with respect to both its detection efficiency and its response behavior [38, 39]. In conventional reference-free XRF geometry (45° incident angle), one may reach an uncertainty of 0.7 % for the solid angle of detection due to a calibrated diaphragm placed at a well-defined distance [38, 39]. However, under grazing incidence conditions (GIXRF) the solid angle of detection can be determined only with an uncertainty of 4.0 % [40]. In Figure 2, a schematic drawing of the experimental setup used for the reference-free XRF and GIXRF is shown.

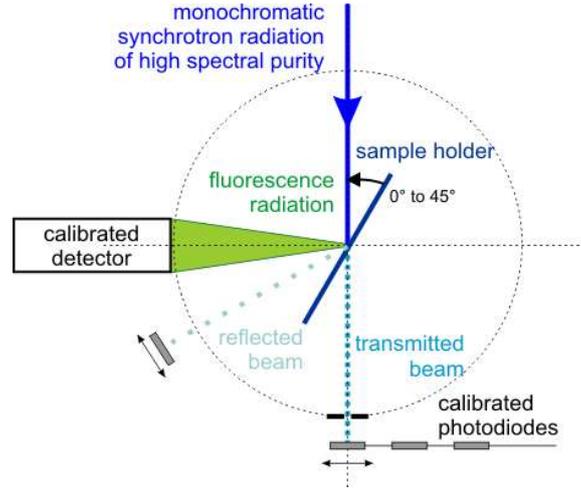

*Figure 2: Scheme of the experimental setup used for the measurements of the X-ray fluorescence radiation in various beam geometries. The direct beam, the reflected beam and (if the sample allows for transmission of the excitation radiation) the transmitted beam can be detected using calibrated photodiodes.*

The combined method of GIXRF measurement and the calculation of the intensity distribution of the XSW field is described in detail elsewhere [41] and will just be summarized here. GIXRF measurements are basically a variation of the incident angle around the critical angle of total external reflection while the element-specific fluorescence radiation is detected. A XSW field is formed due to interference of the incident and reflected beam. The intensity of the XSW field is strongly dependent on the incident angle and modifies the fluorescence line intensity emitted by atoms inside the XSW field. The shape of the curve of the fluorescence line intensity in an XSW field as a function of the incident angle varies for different objects such as nanoparticles, thin films or implantation profiles [42].

### 2.3. Reference-free XRF quantification

For the reference-free quantification, knowledge of all relevant instrumental and experimental parameter is needed as well as the respective fundamental parameters [43, 44, 45, 46]. The mass deposition $m_i/F_I$ of the element $i$ with the unit area $F_I$ can be calculated using the following equation [15, 40, 47]:

$$\frac{m_i}{F_I} = \frac{-1}{\mu_{tot,i}} \ln\left\{1 - \frac{P_{i,j}}{P_0 \cdot I_{XSW(E_0)} \cdot Q(E_0) \cdot \frac{\Omega_{det}}{4\pi} \cdot \frac{1}{\sin\Psi_{in}} \cdot \frac{1}{\mu_{tot,i}}}\right\},$$

with $\mu_{tot,i} = \frac{\mu_{i,E_0}}{\sin\Psi_{in}} + \frac{\mu_{i,E_{i,j}}}{\sin\Psi_{out}}$ and $Q(E_0) = \tau_{X_i,E_0} \cdot \varpi_{X_i} \cdot g_{j,X_i}$.

Here, $\mu_{i,E_0}$ is the mass attenuation coefficient for the element $i$ at the excitation photon energy $E_0$,

$\mu_{i,E_{i,j}}$ is the mass attenuation coefficient for the element $i$ at the photon energy $E_{i,j}$, $\tau_{X_i,E_0}$ is the partial photoelectric cross section of the absorption edge $X_i$ (of the element $i$), $\varpi_{X_i}$ is the fluorescence yield of the absorption edge $X_i$ (of the element $i$) and $g_{j,X_i}$ is the transition probability of the fluorescence line $j$.

$P_{i,j}$ is the photon flux from the fluorescence line $j$ of the element $i$, $P_0$ is the photon flux of the incident radiation, $I_{XSW(E_0)}$ is the relative intensity of the XSW field, $\Omega_{det}$ is the solid angle of detection, $\Psi_{in}$ is the incident angle and $\Psi_{out} = 90° - \Psi_{in}$ the observation angle.

### 2.4. NEXAFS investigations

The NEXAFS measurements at the Ti $L_{3,2}$ absorption edges were carried out at the PGM beamline. Prior to the NEXAFS experiments, angular dependent GIXRF measurements were conducted to determine the incidence angle for the NEXAFS experiments, as exhibited in Figure 7 (right). These curve shapes suggested using an incidence angle of 2.44° for the measurements, because this incident angle is below the critical angles of total external reflection and, therefore, the excitation intensity is modified by the XSW field.

The NEXAFS spectra were recorded in fluorescence detection mode, which means that for every photon energy a fluorescence spectrum was acquired by an energy-dispersive SDD and the Ti L fluorescence line intensities were evaluated. Afterwards, they were normalized to the incident photon flux and the efficiency of the detector. In this work, no further background corrections are necessary due to the excitation conditions chosen.

## 3. Results

In this work, the characterization of the core-shell nanoparticles consists of the determination of the elemental composition, the surface coverage, which is the main part of the characterization, and the chemical speciation, i.e. information on the binding state. For the determination of the surface coverage, the reference-free, fundamental parameter-based XRF analysis is used to determine the mass deposition of titanium.

The knowledge of the mass deposition of titanium then leads to the determination of the number of Pt-Ti core-shell nanoparticles on the surface using the a-priori information about the weight or specified single particle mass (30000 u) of the core-shell nanoparticles, and in the final step to the surface coverage with the assumption that the entire mass lies within the unit area of 1 cm$^2$. In order to do that, the areal weight (mass) ratio from titanium and platinum has to be determined as well, which is described in the next section. This knowledge of the surface contamination or surface elemental mass deposition is then used for the investigation of the XSW field calculation with regard to its degradation.

### 3.1. Determination of the elemental composition

In order to link dimensional nanometrology with analytical nanometrology, the areal weight ratio of titanium and platinum must be known. The areal weight ratio is the ratio of the mass deposition between Ti and Pt. Therefore, the elemental composition of the core-shell nanoparticles is analyzed using reference-free GIXRF. The excitation energy was set to 12.5 keV, which is above the titanium K- and the platinum L3 edges. The incident angle was set to 0.125° with respect to the sample surface, which is below the critical angle of total external reflection, in order to have a decent signal-to-background ratio. In Figure 3, the corresponding fluorescence spectrum is shown for a sample with a nominal surface coverage of 5.0 %.

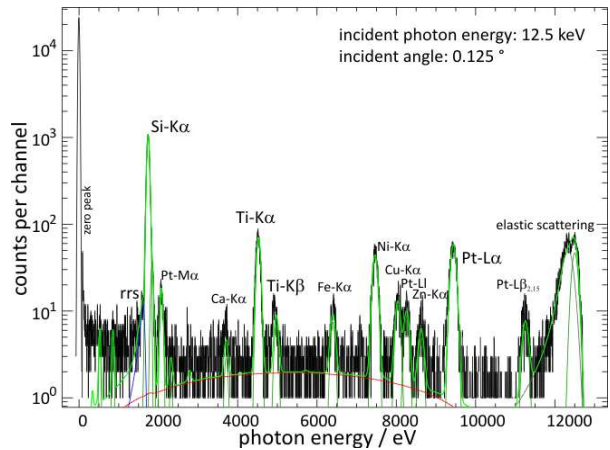

*Figure 3: GIXRF spectrum of a sample with nominal 5 % surface coverage as an example of the determination of the elemental composition. Titanium and platinum can be identified.*

Figure 3 depicts a typical GIXRF spectrum. The titanium Kα and Kβ fluorescence lines are clearly visible as well as the platinum Lα fluorescence line, not only confirming the elements titanium and platinum as elemental components of the sample but also allowing for a reliable quantification. The silicon signal from the substrate is also clearly visible in the spectrum. The O fluorescence line originates from the 1.5 nm thick native $SiO_x$ layer of the substrates but cannot be seen in the depicted spectrum due to the high excitation energy and the background of the substrate fluorescence intensity. Minor contaminations by the elements Ca, Fe, Ni, Cu and Zn are visible. It might be caused by the production, cleaning or transportation process. Ni might be a substrate contamination. The quantification with reference-free XRF of both elements leads to a Ti:Pt areal weight (mass) ratio of 0.63(5) : 0.37(5) as against the nominal 0.75 : 0.25 sputter target ratio. The information on the mass ratio is independent of the absolute mass deposition, which is why an even simpler approach can be used for the determination of the surface coverage (see respective section 3.2 below). The instrumental and experimental parameters such as the solid angle of detection and the incident photon flux are the same for both quantifications and therefore do not contribute to the uncertainty budget in this case. The main contributions to the uncertainty budget are the relevant fundamental parameters.

*3.2. Determination of the elemental mass deposition and homogeneity*

The varying intensity distribution of the XSW-field influences the fluorescence line intensity of an element for different incident angles. Therefore, knowledge about the XSW-field intensity is necessary for a reliable quantitative analysis. But, the surface coverage of the core-shell nanoparticle can modify the XSW-field and thereby change the observed fluorescence intensities. Therefore, a determination of the surface coverage is necessary for a further analysis of the XSW-field.

A detailed characterization of the core-shell nanoparticle which reveals the surface coverage can be performed using the combination of the measured fluorescence intensity of an element in the sample (here the titanium Lα,β fluorescence lines), as a function of the incident angle of the excitation radiation (GIXRF), and a calculation of the X-ray standing wave (XSW) field.

The parameters needed for the calculation of the XSW field provide additional information about the core-shell nanoparticle, such as surface coverage. The GIXRF measurements were performed in the soft X-ray range, because the critical angle of total reflection at the substrate is considerably higher in this regime. As a result, the relevant angular regime for the fluorescence line intensity in the soft X-ray range is shifted and stilted towards higher incident angles in comparison to hard X-rays. This leads to a reduction in the uncertainty of the solid angle of detection in the relevant angular regime. Using the reference-free XRF quantification algorithm [22, 23, 40] way above the critical angle of total external reflection, the relative XSW field intensity changes to unity and is no longer relevant for the quantification of the mass deposition.

In the soft X-ray range, the total uncertainty of the reference-free quantification is dominated by the uncertainty contribution of the fundamental parameters in this energy range, which can be up to 40% [44]. In order to reduce the total uncertainty of the quantification in the soft X-ray regime, samples with nominal 7.5 % and 15 % surface coverage were also analyzed in the hard X-ray range with an incident photon energy of 5.5 keV, where the fundamental parameters used have lower uncertainties. The titanium K$\alpha$ fluorescence line intensity is shown as a function of the incident angle of the excitation radiation in Figure 4.

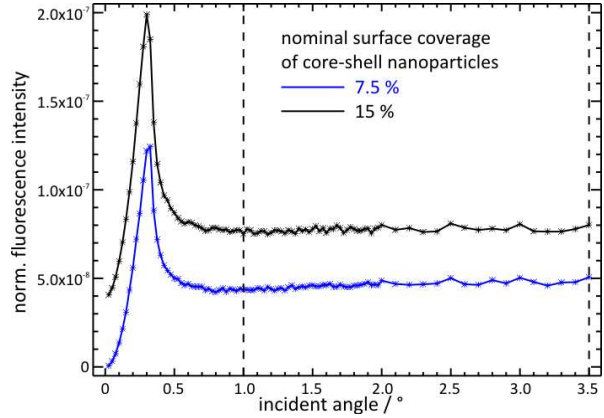

*Figure 4: GIXRF measurement in the hard X-ray range (titanium K$\alpha$ line) of samples with nominal 7.5 % and 15 % surface coverages. The dotted lines indicated the range without XSW-field contribution for the quantification.*

For the quantification, the incident angle of the excitation radiation was chosen to be above the critical angle of total external reflection for the silicon substrate in order to have no XSW field contribution in the excitation condition, indicated by the dotted lines in Figure 4. The mean values of the determined mass deposition of titanium in the hard X-ray range are:

$$\frac{m_{Ti,7.5\%}}{F_I} = 4.6x10^{-8} \, g/cm^2 \pm 4.0x10^{-9} \, g/cm^2$$

$$\frac{m_{Ti,15\%}}{F_I} = 7.7x10^{-8} \, g/cm^2 \pm 6.8x10^{-9} \, g/cm^2$$

The uncertainty budget with a relative uncertainty of about 8.7 % is still dominated by the respective fundamental parameters, but is much lower than the comparable uncertainties in the soft X-ray range. The uncertainty budget is presented in section 5 below. The quantification of the mass deposition of titanium for the samples with nominal 7.5 % and 15 % surface coverages in the hard X-ray range can be used for the determination of a fluorescence production cross-section for the quantification in the soft X-ray range of the same samples.

The incident photon energy in the soft X-ray range was chosen to be 520 eV, which is above the titanium L2-edge. As a result, the titanium L3 and L2 lines appear in the spectrum and cannot be separated by the SDD used. The quantification with this excitation condition also involves more fundamental parameters such as the L3 and L2 subshell photoionization cross sections and the L2-L3 Coster-Kronig transition probability, which increases the uncertainty of the determined mass deposition. The determined mass deposition of titanium in the hard X-ray range can be used for the determination of the fluorescence production cross-section in the soft X-ray range of the same samples. This fluorescence production cross-section can then be used for the determination of the mass deposition of titanium in all other samples in the soft X-ray range. The uncertainty of the determined mass deposition in the soft X-ray range is significantly reduced by using this fluorescence production cross-section.

However, the inhomogeneity of the sample used for the correction factor has to be rather small in order to minimize the uncertainty of this correction. The fluorescence production cross-section has been determined for the two samples, and they are not identical due to the inhomogeneity of the samples. Here, the mean value of the two determined correction factors is used for the quantification in the soft X-ray range of all samples. The inhomogeneity of the samples dominates the total uncertainty budget for the correction of the quantification in the soft X-ray range. In order to reduce this contribution, the lateral distribution of the core-shell nanoparticles could be mapped with a $\mu m$-sized synchrotron beam [49] and considered in the correction. In Table 1, the quantified titanium mass depositions for each sample are listed.

| Sample / nominal surface coverage | Titanium mass deposition† / g / cm² | Uncertainty / g / cm² (k=1) |
|---|---|---|
| 2.5 % | $4.4 \times 10^{-8}$ | $9.6 \times 10^{-9}$ |
| 5.0 % | $5.3 \times 10^{-8}$ | $1.2 \times 10^{-8}$ |
| 7.5 % | $5.3 \times 10^{-8}$ | $1.2 \times 10^{-8}$ |
| 15 % | $6.8 \times 10^{-8}$ | $1.5 \times 10^{-8}$ |
| 30 % | $5.0 \times 10^{-7}$ | $1.1 \times 10^{-7}$ |
| 60 % | $8.3 \times 10^{-7}$ | $1.8 \times 10^{-7}$ |

*Table 1: Determined mass deposition of titanium for core-shell nanoparticle samples with different surface coverage.*

As shown in Figure 5 (left hand side), the quantified mass deposition of titanium is not linearly increasing with the nominal surface coverage. On the right-hand side of Figure 5, for a comparison with the nominal total surface coverage, the peak surface coverage was determined and used for the x-coordinate.

---

† The soft x-ray quantification was performed using the fluorescence production cross-section derived from the absolute quantification in the hard X-ray range.

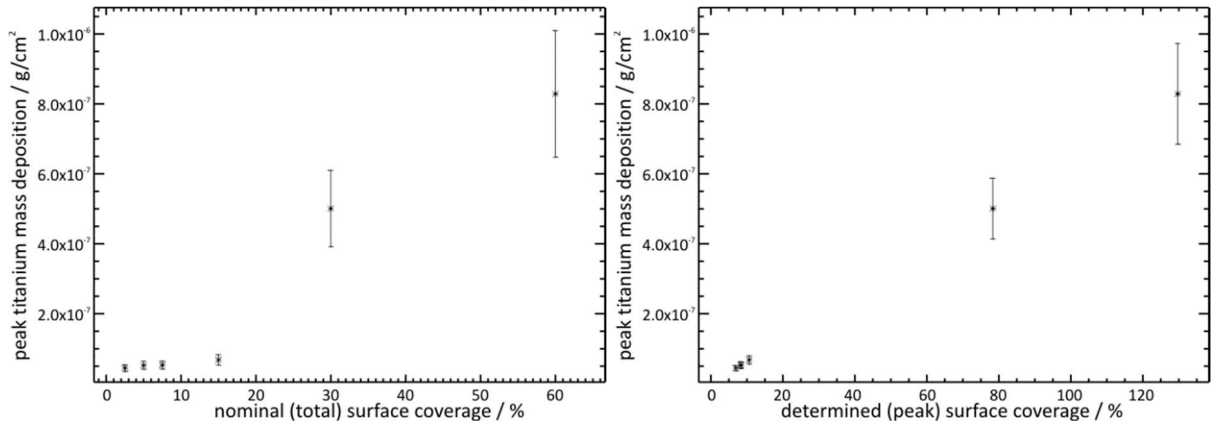

*Figure 5: Titanium mass deposition for the core-shell nanoparticle sample as a function of the nominal and determined (peak) surface coverage.*

However, the quantification has been performed at the center or peak position of each sample, allowing us to derive a peak mass deposition. The peak surface coverage is the value, which is calculated using the determined mass deposition of titanium at the peak position of the lateral scan with the assumption that the determined mass deposition is homogeneously distributed along the entire sample. In order to classify the quantification, the homogeneity of the samples has to be considered. Therefore, measurements of the lateral homogeneity have been performed for all samples. The deviation of the mass deposition is up to 60 % and the deposition is not centered on all samples. That can explain the deviation of the linear correlation in the mass deposition of the different sample surface coverages. Figure 6 shows the lateral homogeneity of all samples.

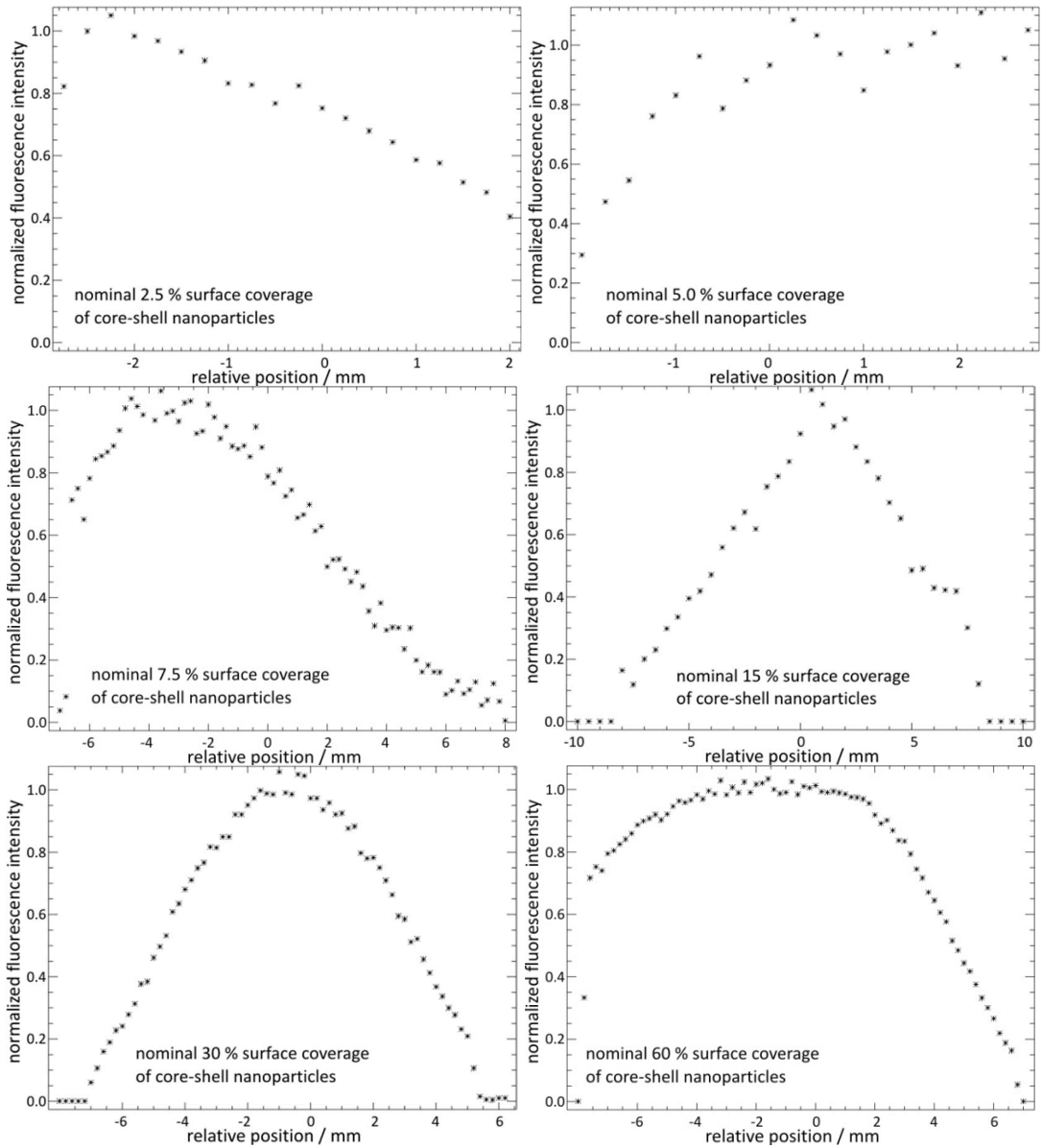

*Figure 6: In order to classify the quantification, the homogeneity of the samples has to be considered. Therefore, measurements of the lateral homogeneity have been performed for all samples. The variation in the intensity is up to 60 % with respect to the central deposition area.*

### *3.3. Determination of the binding state*

In Figure 7 the Ti L3,2 NEXAFS spectra of the samples with a nominal surface coverage of 2.5% and 5%, for an incidence angle below the critical angle of total reflection are exhibited. The curves are very similar and suggest the same chemical binding state. There are pronounced peaks at the L3 absorption edge at 458 eV, 459.6 eV, and 460.4 eV, which arise due to splitting caused by the crystal field surrounding the O atoms [50]. This behavior can be observed at the L2 edge as well, but these peaks are broadened due to the shorter lifetime [51]. From these findings we conclude that the shell of the particles solely consists of amorphous titanium dioxide. No other Ti species contributions are observable, thus allowing to state that the particle size is not varying due to changes in the structure or related species.

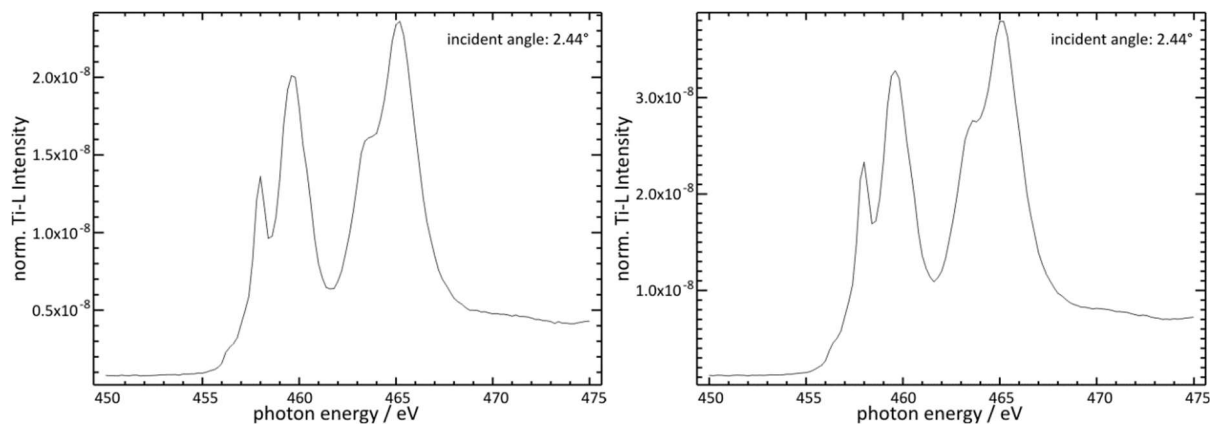

*Figure 7: Ti L3,2 NEXAFS measurements of samples with nominal 2.5 % and 5 % surface coverages. The incidence angle was chosen to be below the critical angle of total reflection in order to receive information about the core-shell nanoparticles, located at the surface of the substrate. The shapes of the curves are similar and suggest the same chemical binding state, titanium dioxide.*

This confirms that the titanium is completely oxidized to $TiO_2$, thus determining the binding state between Ti and Pt. The change of the surface coverage has no impact on the NEXAFS spectra.

## 4. Discussion

### 4.1. Link of dimensional and analytical nano-metrology

As shown in 3.1, the areal weight (mass) ratio of Pt and Ti was determined with reference-free XRF to 0.63(5) : 0.37(5) as against the nominal 0.75 : 0.25 sputter target ratio.

The knowledge of the total mass of Pt and Ti in one core-shell nanoparticle in combination with the determined mass ratio allows for an estimation of the average size of a core-shell nanoparticle. The bond lengths of 277.5 pm for the Pt-Pt bond and 162.0 pm for the Ti-O bond were taken from Sutton et al. [48] to calculate the volume of the respective atoms. For Pt, the number of atoms is determined to be about 57 per cluster and for Ti, about 395 per cluster or single core-shell nanoparticle, respectively. The core-volume consists of Pt and the shell-volume of Ti in this estimation, amounting to a size of about 1.6 nm in diameter per cluster. Here, the close-packing of equal spheres was estimated for the calculation of the volume (packing density about 74 %).

That is lower than the result of the analysis of the STEM images, which is 2 nm to 3 nm. A possible explanation is that the current estimation is to calculate the diameter of a sphere, not considering any deviating arrangement of the nanoparticles on a surface area, which would increase the diameter significantly. For an estimation with the assumption that all Pt and Ti atoms are arranged in one layer (circle-area), the diameter of one core-shell nanoparticle increases to about 3.8 nm. So the STEM value-interval lies well in between the sphere size estimation and the circle-area size estimation. Other explanations may be that the estimated bond lengths and the estimation for the packing density do not match the reality well enough.

### *4.2. Estimation of the surface coverage*

An estimate of the surface coverage using the quantification of the titanium mass deposition can be done as follows: The assumption of particles not laying on top of each other in conjunction with the a-priori information about the mass and size of one core-shell nanoparticle leads to the number of particles in the area of one square-centimeter (1 cm$^2$). One core-shell nanoparticle has the nominal size of 2 nm to 3 nm in diameter (determined with STEM), so the area of 1 cm$^2$ can contain about $2 \times 10^{13}$ nanoparticles to reach 100 % surface coverage. For this estimation, the average size of the core-shell nanoparticles was assumed to be 2.5 nm and the area of one core-shell nanoparticle was assumed to be circular. The mass of one core-shell nanoparticle is specified to 30000 u (Pt/Ti) before oxidation and the compounds are specified to Pt/TiO$_2$, as confirmed using NEXAFS measurements (see section above). The oxidation of titanium increases the total mass of the core-shell nanoparticle, but is not considered in the 30000 u. The mass proportion for titanium and platinum were experimentally determined to be 0.63(5) and 0.37(5). The mass of oxygen in TiO$_2$ is not considered in the estimation, as the specified weight of the core-shell nanoparticles did not include the oxidation of titanium. This leads to a fraction of titanium mass in the core-shell nanoparticle of about 63 %. Thus, the mass of titanium in one core-shell nanoparticle is about $3.1 \times 10^{-20}$ g. The determined mass depositions of Ti for the different samples, as shown in table 1, lead to the number of core-shell nanoparticles. The number of core-shell nanoparticles determined using the mass deposition of titanium and the fraction to 100 % surface coverage are listed in Table 2. The uncertainty consists of the contribution of the reference-free quantification, the mass per core-shell nanoparticle and the mass ratio of Ti and Pt.

| Sample / nominal surface coverage | determined number of core-shell nanoparticles (peak) | determined peak or probed surface coverage |
|---|---|---|
| 2.5 % | $1.41 \times 10^{12} \pm 3.1 \times 10^{11}$ | 7 % $\pm$ 1.6 % |
| 5.0 % | $1.69 \times 10^{12} \pm 3.8 \times 10^{11}$ | 8 % $\pm$ 2.0 % |
| 7.5 % | $1.70 \times 10^{12} \pm 3.8 \times 10^{11}$ | 8 % $\pm$ 2.0 % |
| 15 % | $2.17 \times 10^{12} \pm 4.8 \times 10^{11}$ | 10 % $\pm$ 2.5 % |
| 30 % | $1.60 \times 10^{13} \pm 3.6 \times 10^{12}$ | 78 % $\pm$ 18.6 % |
| 60 % | $2.64 \times 10^{13} \pm 5.9 \times 10^{12}$ | 130 % $\pm$ 30.8 % |

*Table 2: Determined amount of core-shell nanoparticles using the determined mass deposition of titanium for all samples and calculated surface coverages of the unit area using the specified mass per nanoparticle of 30000 u. The uncertainty of the surface coverage consists of the contribution of the reference-free quantification, the mass per core-shell nanoparticle and the mass ratio of Ti and Pt.*

It can be seen that the results for the determined peak surface coverages vary, but are often too high in comparison with the nominal total values. In Figure 5 (right-hand side), the determined mass deposition of titanium and the respective surface coverage are shown for each samples. One reason for this result can be the estimation of the nominal size and mass of a core-shell nanoparticle for the calculation of 100 % surface coverage. Another reason can be the lateral inhomogeneity of the deposited nanoparticles. The estimation does not consider a non-homogeneous distribution of the nanoparticles, which always overestimates the surface coverage when measuring at the position with the most material. Especially the samples with nominal surface coverage of 30 % and 60 % seem to have a Gaussian distribution of the core-shell nanoparticle on the sample. Measurements at the center position lead to an overestimation of the total mass deposition by a factor of about 2, assuming that all particles are in the unit area of 1 cm². Taking this effect into account, the effective mass deposition is reduced, and thereby the related surface coverage reduces to 39 % and 65 %, respectively, thus better allowing to link nominal values with experimentally derived values of the surface coverage rate.

### *4.3. Modification of the X-ray standing wave field*

The surface coverage of the Pt-Ti core-shell nanoparticles also influences the intensity distribution within the XSW field. For a very low surface coverage, the XSW field above the surface of the reflecting medium is undisturbed, and can in the present case, be calculated using the optical constants of the silicon substrate and vacuum above the reflecting surface. With increasing coverage, the reflection properties of the surface change due to the nanoparticles and have to be taken into account in the XSW field calculation. Using an effective density approach, where the nanoparticles are assumed to be a continuous layer with reduced density to take into account the coverage, this can be done. Such calculations, which were performed using software developed in-house [24] are shown in Figure 8 (left-hand side) in comparison with the experimental Ti-L$\alpha$,$\beta$ fluorescence GIXRF measurements. Here, for the samples with nominal 2.5 %, 5.0 %, 30 % and 60% surface coverage, the relative intensity of the peak at the critical angle of total external reflection is clearly decreasing with increasing coverage, whereas the angular position of the peak does not change significantly. All GIXRF measurements have been performed at the peak position of the respective sample deposition. The same behavior is found for the calculated curves, assuming a thin layer with varying density. This clearly demonstrates that the particles must be taken into account during the XSW field calculation. The samples with nominal 7.5 % and 15 % surface coverages have a larger nitrogen contamination than the other samples, which influences the Ti-L$\alpha$,$\beta$ fluorescence intensity distribution in the GIXRF measurements, and are not included here for this reason.

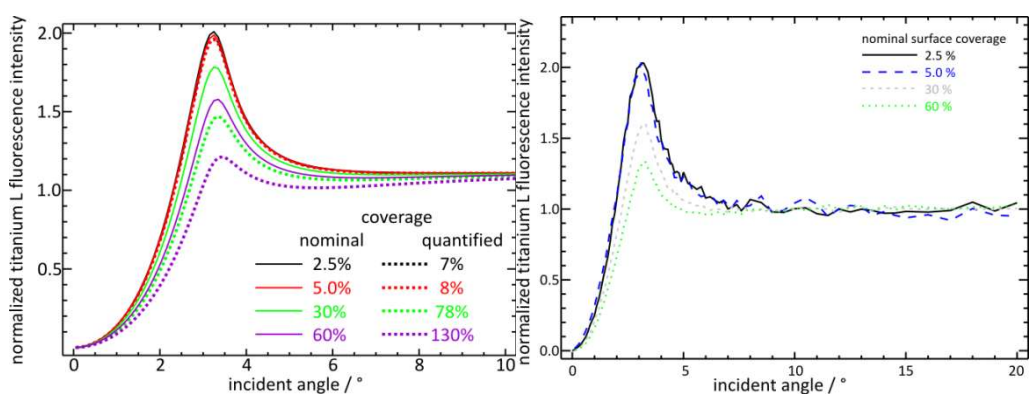

*Figure 8: Calculated relative XSW field intensity (left) and GIXRF measurements in the soft X-ray range (titanium Lα,β fluorescence lines) of all samples with different nominal surface coverage (right). The shape of the curve indicates different intensities of the XSW field. At high incident angles, the excitation radiation is penetrating into the sample and is not reflected anymore, so that no relevant XSW contribution occurs.*

## 5. Uncertainty budget

For the quantification in the hard X-ray regime, the uncertainties of the fundamental parameters are the main contribution to the total uncertainty budget. The correction factor for the quantification in the soft X-ray range has a major contribution to the uncertainty due to the inhomogeneity of the sample used. In Table 3, all contributions to the uncertainty budget are listed.

| parameter | relative uncertainty ($10^{-2}$) | Comment |
|---|---|---|
| $S_0$ | 0.01 | Signal of the photodiode [29] |
| $\sigma_{diode,E_0}$ | 1.0 | Spectral response of the photodiode [29] |
| $\Phi_0(E_0)$ | 1.0 | $\Phi_0(E_0) = S_0/\sigma_{diode,E_0}$ |
| $R^d_{i,line}$ | 2.0 | Spectral deconvolution: statistics & overlaps |
| $\varepsilon_{det,E_i}$ | 1.5 | SDD calibration [39, 52] |
| $\Phi^d_{i,line}$ | 2.5 | $\Phi^d_{i,line} = \dfrac{R^d_{i,line}}{\varepsilon_{det,Ei}}$ |
| $\Omega_{det}$ | 4.0 | Solid angle of detection [9, 40] |
| $\omega_K$ | 5.0 | K fluorescence yield of titanium |
| $\mu_{Ti}(E_0, E_i)$ | 2.0 | Mass attenuation coefficient for $E_0$ and $E_i$ [43] |
| $\tau_{Ti,K}(E_0)$ | 5.0 | K-subshell photoionization cross-section for $E_0$ [43] |
| **I** | 8.7 | Total uncertainty for the mass deposition using K-Fluorescence |
| **II** | 20.0 | Fluorescence production cross-section correction factor |
| Total (L-Fluorescence) **III** | 21.8 | Quadrature sum of {**I** and **II**} |

| Mass of one nanoparticle **IV** | 5.0 | contributes to the uncertainty of the surface coverage |
|---|---|---|
| Mass ratio Ti : Pt {**V**} | 8.0 | contributes to the uncertainty of the surface coverage |
| Total (surface coverage) | 23.8 | Quadrature sum of {**III, IV** and **V**} |

*Table 3: The relative uncertainties of the quantification in GIXRF geometry above the critical angle of external total reflection are shown.*

# 6. Summary, conclusion and perspectives

In this work, two complementary investigations were performed: On the one hand side, very monodisperse core-shell nanoparticles of titanium and platinum were analyzed by means of Si traceable XRF. On the other hand side, those characterized particles were employed under GIXRF conditions to probe the modification of the XSW-field intensity due to surface coverage by these core-shell nanoparticles. The reference-free GIXRF analysis using hard X-ray radiation for the excitation leads to a mass ratio of titanium 0.63(5) and platinum 0.37(5). This allows for a good estimation of the particle size, using the specified single particle mass of the core-shell nanoparticles as a-priori information and a close packing of their atoms as assumption. Thereby, analytical metrology and dimensional nanometrology could be combined. This interaction of NP and X-ray analytical techniques has the result that core-shell nanoparticles modify the intensity distribution of the XSW field with increasing surface coverage. On the other hand, a characterization of an unkonwn mass deposition of NP has to take this effect into account. For the quantitative analysis, the titanium K lines were used for a fluorescence production cross section correction factor in the soft X-ray range in order to reduce the uncertainties. The determined mass deposition of titanium for the samples with different nominal core-shell nanoparticles surface coverage ranges from $4.4 \times 10^{-8}$ g / cm$^2$ to $8.3 \times 10^{-7}$ g / cm$^2$. This leads to a surface coverage of the core-shell nanoparticles, based on the composition determination above and with the assumption of a specified single particle mass of 30000 u, from 7 % to 130 %. The uncertainty of the surface coverage is dominated by the lateral inhomogeneity of the core-shell NP deposition on the substrate for two reasons: First, it increases the uncertainty of the fluorescence production cross-section correction factor being used to ensure sufficient soft X-ray sensitivity. Second, the surface coverage concept is based on the assumption of a homogeneously distributed mass deposition and any deviation from this cause the respective uncertainty to increase. The nanoparticles' surface coverage influences the XSW field intensity for incident angles in the region of the critical angle of external total reflection, which is observable in the shape of the fluorescence line intensity of titanium L lines for different surface coverages. A calculation of the angular dependence of the XSW field intensity can simulate the shape of the fluorescence line intensity and shows the strength of this method. The calculated XSW field intensity follows the shape of the experimental determined fluorescence line intensity for the core-shell nanoparticles used here. For nanoparticles way larger than 3 nm, this kind of behavior has to be confirmed. A further step, in order to improve the XSW based quantification, is to perform an iterative forward calculation of the XSW field intensity with the surface coverage of the core-shell nanoparticles as a fitting parameter. The surface coverage as a fitting parameter is independent of any a-priori information on the nanoparticle weight.

# Acknowledgements

The study regarding the characterization of core-shell nanoparticles was related to the EMPIR project 14IND12 "Innanopart" and 16ENV07 "Aeromet". The financial support of the EMPIR program is gratefully acknowledged. It is jointly funded by the European Metrology Programme for Innovation and Research (EMPIR) and participating countries within the European Association of National Metrology Institutes (EURAMET) and the European Union.